\documentclass[a4paper]{article}

\usepackage{amsmath,amssymb,amsfonts,amsthm} 
\usepackage{graphicx}

\usepackage{verbatim}

\usepackage[margin=1cm,font=small]{caption}

\usepackage{layout} 

\numberwithin{equation}{section}

\input{Macros.tex}

\newcommand{\boldvarsigma}{{\boldsymbol{\varsigma}}}
\newcommand{\bolde}{{\mathbf{e}}}
\newcommand{\boldrr}{{\mathbf{r}}}
\newcommand{\boldq}{{\mathbf{q}}}
\newcommand{\boldx}{{\mathbf{x}}}
\newcommand{\boldr}{{\mathbf{r}}}

\begin{document}

\title{Energy fluctuations in simple conduction models}
\author{
Fran\c cois Huveneers\footnote{
CEREMADE, 
Universit\' e de Paris-Dauphine,
Place du Mar\'echal De Lattre De Tassigny,
75775 PARIS CEDEX 16 - FRANCE.
E-mail: huveneers@ceremade.dauphine.fr.
Supported by the European Advanced Grant Macroscopic Laws and Dynamical Systems (MALADY) (ERC AdG 246953)}
}

\maketitle

\begin{abstract}
\noindent
We introduce a class of stochastic weakly coupled map lattices, as models for studying heat conduction in solids. 
Each particle on the lattice evolves according to an internal dynamics that depends on its energy, 
and exchanges energy with its neighbors at a rate that depends on its internal state. 
We study energy fluctuations at equilibrium in a diffusive scaling.
In some cases, we derive the hydrodynamic limit of the fluctuation field. 
\end{abstract}

\vspace{10cm}

\pagebreak


\section{Introduction}\label{sec: Introduction}

It is generally expected that the temperature profile of thermally isolated solids relaxes according to some non-linear heat equation
\begin{equation*}
\partial_t T \; = \; \nabla \big( D(T) \nabla T \big).
\end{equation*}
The derivation of this equation, starting from a microscopic hamiltonian dynamics, is one of the major open problems in statistical mechanics out of equilibrium \cite{bon}\cite{liv}.
Very idealized models of solid constitute a possible starting point to develop our understanding on the problem, and in fact,
at the present time, even such cases can be challenging to analyze
(see \cite{eck} for a good example of such a philosophy).

Lattice gases furnish the first examples where the heat equation can be recovered through a diffusive rescaling of space and time.
Each atom reduces here to a point on a lattice, is characterized by its energy only, 
and interacts stochastically with its neighbors.
The stochastic interaction is a reasonable approximation for particles that should evolve according to some chaotic internal dynamics. 
A set of independent random walks (IRW), the simple symmetric exclusion process (SSEP) \cite{kip}, and the Kipnis Marchioro Presutti model (KMP) \cite{kip2},
constitute three cases where the heat equation can be easily derived.  
These systems may be closer to physics than what could be thought at the first glance: 
in \cite{huv}, an example of noisy hamiltonian system is found, that gives rise to the SSEP in a weak coupling limit.

This said, true atoms should clearly be described by position and momentum, and not only energy. 
To move towards a more realistic situation, 
let us thus consider a lattice where particles at the nodes now have some internal degree of freedom.  
We assume the local dynamics of particles to have good ergodic properties, and the interaction between them to be weak and controlled by a parameter $\epsilon >0$.
So, individual atoms will reach their own equilibrium at a smaller time scale than energy is exchanged.

Much progress has been recently accomplished in the understanding of these systems, 
by means of a two-stage analysis, as outlined in \cite{gas}.
First, letting $\epsilon \rightarrow 0$ and rescaling time properly, an autonomous stochastic process for the energy alone is obtained.
Second, rescaling space and time diffusively, the heat equation is derived starting from this mesoscopic stochastic process. 
The first part of this program has now been rigorously accomplished for a purely hamiltonian dynamics \cite{dol}.
Though in a rather different way, the distinction of two different time-scales also has allowed to better understand the conductivity of weakly anharmonic solids \cite{aok}\cite{luk}.

The introduction of two successive limits is however somehow artificial. 
One eventually would like to fix some, possibly small, $\epsilon > 0$, and directly consider the diffusive limit.   
Results in that direction are few. 
Let us quote three of them.
First, in \cite{ber}, the heat equation is recovered starting from a noisy harmonic chain (see also \cite{fri} for fluctuations at equilibrium). 
The proof rests on the fact that an exact fluctuation-dissipation decomposition is available there.
Second, a deterministic coupled map lattice is analyzed in \cite{bri}, 
and diffusion of an initially localized energy packet is shown. 
The dynamics of particles is there assumed to be independent of energy. 
Third, fluctuations of energy for a noisy hamiltonian system at equilibrium are considered in \cite{oll} (see also \cite{her} for a simpler model) ; 
the fluctuation field is shown to converge to a generalized Ornstein-Uhlenbeck process.
Some level of symmetry of the generator, the sector condition, is needed to derive this result.

In this paper, we introduce models of heat conduction that are conceptually very simple. 
They maybe will even look a bit cheap compared to the aforementioned examples.
Though, the analysis reveals that nowadays techniques do not directly apply to describe how energy diffuses in the cases we consider. 
We will in fact need to look at particular situations in order to derive rigorous results. 
To make this more concrete, let us now briefly outline the content.

In Section \ref{sec: Models}, we define three models of conduction.
In a sense made precise by \eqref{projection energy space} below, 
they constitute respective versions of the IRW, SSEP and KMP models, where, now, atoms have an internal degree of freedom. 
The interaction between near particles is controlled by a parameter $\epsilon >0$. 
The models so obtained are non-gradient and do not satisfy the sector condition.

Like in \cite{oll} and \cite{her}, we study the fluctuations of energy in equilibrium, as this might be the easiest problem to look at.
We introduce the fluctuation field in Section \ref{sec: Fluctuation field}, 
and we state there Proposition \ref{the: limit fluctuation field}, that will allow us to derive the hydrodynamic limit of this field in some cases.
While it is essentially a variation on a classical method \cite{kip}, 
we here assume that exists an approximative fluctuation-dissipation decomposition of the microscopic current in a particular sense. 
We hope that this kind of result can be of some use elsewhere. 
The proof is postponed to Section \ref{sec: proof limit fluctuation field}.

Explicit results are gathered in Section \ref{sec: Explicit results}.
The hydrodynamic limit of the field in the first order in $\epsilon$ is derived in Theorem \ref{the: first order}.
As a corollary, a weak coupling limit is shown, 
where $\epsilon$ is sent to 0 after a diffusive rescaling  of space and time. 
This kind of limit is a bit more satisfactory than the one in \cite{dol}\cite{huv}\cite{liv}, where $\epsilon$ was sent to 0 for a fixed number of particles without any rescaling of space.
It also avoids the two-stage procedure of \cite{gas}.
One then approximates, like in \cite{bri}, the local dynamics by a process independent of energy. 
In Theorem \ref{the: independent dynamics}, the hydrodynamic limit of the fluctuation field is obtained under that hypothesis for any small enough $\epsilon >0$.
While, with respect to \cite{bri}, we deal with a rather particular case, our proof is much shorter, and we are not limited to an initially localized energy packet.

\section{Models}\label{sec: Models}

Throughout all the paper, $\epsilon > 0$ represents a strictly positive number.
Let us consider a set of $N \ge 3$ identical particles.
The coordinates of the particles are indexed by a point in the periodic one-dimensional lattice $\Z_N$ of integers modulo $N$.
They form thus a one-dimensional chain with periodic boundary conditions. 
Each particle is characterized by a local degree of freedom, 
and a non negative energy, so that the phase space $\mathsf X$ is given by 
\begin{equation*}
\mathsf X \; = \; \mathsf M^N \times \R_+^N.
\end{equation*}
where
\begin{equation*}
\mathsf M \; = \; \T \times \{ -1,1\} \; = \; (\R/\Z) \times \{ -1,1\}.
\end{equation*}
A point $r\in \mathsf M$ is written as $r = (q,\varsigma) \in \T \times \{ -1,1\}$ ; 
a point $\boldx\in \mathsf X$ is written as 
\begin{equation*}
\boldx 
\; = \; 
(\boldr , \bolde) 
\; = \; 
(r_1 , \dots , r_N , e_1, \dots , e_N) 
\; = \; 
(\boldq , \boldvarsigma, \bolde) 
\; = \;  
(q_1 , \dots , q_N , \varsigma_1, \dots , \varsigma_N, e_1, \dots , e_N) .
\end{equation*}
We set also $x_k = (r_k,e_k) = (q_k, \varsigma_k, e_k)$ for $1 \le k \le N$.

We see $q_k$ as the position of particle $k$, and $\varsigma_k$ as the sign of its velocity. 
It could have seemed more natural to first define the usual hamiltonian variables $q_k$ and $p_k$, 
and then express the energy in function of them. 
However, for our purposes, the distinction between the local variable $r_k$ and the energy $e_k$ is the most important.
In fact, we almost never will make use of the variables $q_k$ and $\varsigma_k$.
We assume that each particle have a ground state energy $\mathsf e_0 > 0$, 
and we think of $e_k$ as being the extra amount of the energy of particle $k$ with respect to $\mathsf e_0$.

We consider a stochastic dynamics defined through the generator
\begin{equation*}
L \; = \; L' + \epsilon \, L'' \; = \; \sum_{k\in \Z_N} \big( L'_k + \epsilon \, L''_k \big)
\end{equation*}
acting on functions on the phase space $\mathsf X$.
Here $L'_k$ governs the internal dynamics of particle $k$, 
and $L''_k$ the exchanges of energy between particles $k$ and $k+1$.
We denote by $X_t$ the process generated by $L$ at time $t \ge 0$.
Let us now concretely define $L'$ and $L''$.

\subsection{Local dynamics}
Let $1 \le k \le N$. One defines 
\begin{equation*}
L'_k u
\; = \; 
A'_k u + S'_k u 
\; = \; 
\varsigma_k\, \sqrt{\mathsf e_0 + e_k} \multiplication \partial_{q_k} u 
\, + \,
\sqrt{\mathsf e_0 + e_k} \multiplication \big( u(\dots , - \varsigma_k, \dots) - u (\dots, \varsigma_k , \dots) \big).
\end{equation*}
Here $A'_k$ and $S'_k$ denote respectively the symmetric and antisymmetric part of $L'_k$, 
with respect to the uniform probability measure $\mu_\mathrm U$ on $\mathsf M$.  
So particle $k$ moves freely on the circle at speed $\sqrt{\mathsf e_0 + e_k}$ and, at random times, the sign of its velocity gets flipped. 
We will make use of the following notation: given a function $u$ on $\mathsf X$, one writes
\begin{equation*}
\langle u | \mathbf e \rangle \; = \; \int_{\mathsf M^N} u (\mathbf r , \mathbf e) \, \mu_{\mathrm U} (\dd \mathbf r).
\end{equation*}

No sector condition holds. 
Moreover, the dynamics generated by the symmetric part $S_k'$ alone is very degenerated since it does not affect the physical location $q_k$. 
The dynamics generated by $L'$ quickly relaxes to equilibrium:
there exists a constant $c >0$, independent of $N$ such that, 
given a function $u$ satisfying $u(\cdot, \mathbf e) \in \Lp^2(\mathsf M^N, \mu_{\mathrm U})$ and $\langle u | \mathsf e \rangle = 0$ for any $\mathbf e \in \R_+^{N}$, 
it holds that
\begin{equation*}
\| \ed^{L' t} u (\cdot, \mathbf e) \|_{2}
\; \le \;
\ed^{-ct} \| u (\cdot , \mathbf e) \|_{2}.
\end{equation*}
This follows from an explicit diagonalisation of $L_k'$ 
(see also \cite{huv} or \cite{dolb} for two different approaches). 
Given $e_k \in \R_+$, the functions 
\begin{equation*}
\varphi_{l,\tau} (q_k, \varsigma_k) \; = \; \frac{1}{Z(l,\tau)} \big( 2i\pi l \sqrt{\mathsf e_0 + e_k} + \delta_0 (\lambda_{l,\tau}) + \lambda_{l,\tau}\varsigma_k \big) \ed^{2i\pi l q_k},
\end{equation*}
with $Z(l,\tau)$ a normalization factor, and with
\begin{equation*}
 \lambda_{l,\tau} 
\; =  \;
- \sqrt{\mathsf e_0 + e_k} \Big( 1+ i\tau\sqrt{ (2\pi l)^2 - 1} \Big), 
\qquad l \in \Z, \; \; \tau \in \{ - 1, +1 \},
\end{equation*}
form an orthonormal basis of $\Lp^2 (\mathsf M)$ such that 
\begin{equation*}
L_k \varphi_{l,\tau} (q_k, \varsigma_k) \; = \; \lambda_{l,\tau} \varphi_{l,\tau} (q_k, \varsigma_k).
\end{equation*}
It is seen that $\lambda_{l,\tau} = 0$ if and only if $l = 0$ and $\tau = 1$ (with the convention $\sqrt{-1}=+i$), and that otherwise $\Re \lambda_{l,\tau} \le - \sqrt{\mathsf e_0}$.
Our claim then follows by tensorization since all particles evolve independently under the dynamics generated by $L'$.

We will use this information in the following form. 
Let $f$ be a function on $\mathsf X$ such that  $f(\cdot, \mathsf e)\in \Lp^2(\mathsf M^N,\mu_\mathrm U)$ and that  $\langle f|\mathbf e \rangle = 0$ for all $\mathbf e$. 
Then there exists a unique function $u$ on $\mathsf X$ 
such that $u (\cdot , \mathbf e) \in  \Lp^2(\mathsf M^N,\mu_\mathrm U)$ and $\langle u | \mathbf e \rangle = 0$ for all $\mathbf e$, 
that solves the Poisson equation 
\begin{equation}\label{solution Poisson equation L'}
- L' u = f.
\end{equation}
We call this function $u$ the fundamental solution to this Poisson equation. 
Moreover there exists a constant $\mathrm C < +\infty$ independent of $N$ and $f$ such that
\begin{equation}\label{norm solution Poisson equation L'}
\| u (\cdot, \mathbf e) \|_2 \; \le \; \mathrm C \, \|f (\cdot , \mathbf e) \|_2
\end{equation}
for every $\mathbf e\in \R^N$.
This follows from the representation 
\begin{equation*}
u \; = \; \int_0^{\infty} \ed^{L' t} f \, \dd t, 
\end{equation*}
or directly from the explicit diagonalisation of $L'$.

\subsection{Interactions}
In a one-dimensional chain of oscillators evolving according to the laws of classical mechanics, 
the strength of the force between two atoms depends on their relative physical location, 
and the instantaneous current between them is a function their positions in phase space.  
To mimic this feature in our model, let us introduce a nonzero uniformly bounded symmetric function $\chi \ge 0$ on $\mathsf M^2$. 
The function $\chi$ will make the rate of exchange of energy depend on the locations of particles. 
We set also
\begin{equation*}
\overline{\chi} \; = \; \int_{\mathsf M^2} \chi (r_0,r_1) \, \mu_{\mathrm U} (\dd \mathbf r).
\end{equation*}
As an example, one can take
\begin{equation*}
\chi (r_0,r_1) \; = \; \chi (q_0 , q_1) \; = \; \chi_{[a,b]^2} (q_0,q_1), 
\end{equation*}
where $\chi_{[a,b]}$ denotes the characteristic function of some interval $[a,b]\subset \T$. 
Then, the interaction between two particles will only be possible if they both are in the interval $[a,b]$.
 
We will define three possible ways of interacting, and so three different models. 
In the three cases, the interaction will be purely stochastic, 
meaning that the operator $L''$ is symmetric with respect to the canonical invariant measures defined in Section \ref{sec: Fluctuation field}.
This is admittedly in sharp contrast with hamiltonian interactions. 
We may view our models as perturbations of the IRW, SSEP or KMP dynamics respectively. 
Indeed, the operator $\mathcal L$ acting on functions $u$ that depend only on the energies $\mathbf e$, and defined by
\begin{equation}\label{projection energy space}
\mathcal L u \; = \; \frac{1}{\overline{\chi}} \, \langle L'' u | \bolde \rangle ,
\end{equation}
will be nothing else than the generator of these dynamics.

\paragraph{Model 1: Exchanging bits of energy.}
Let us restrict the set of reachable energies to $\N = \{0,1,2, \dots\}$, so that the phase space writes $\mathsf X = \mathsf M^N \times \N^N$.
Let $1 \le k \le N$. 
One defines 
\begin{equation*}
L_k'' u \; = \; 
\chi (r_k , r_{k+1}) \multiplication
\big( e_k (u \circ \sigma_k^{+} - u) + e_{k+1} (u \circ \sigma_{k+1}^{-} - u) \big)
\end{equation*}
with
\begin{align*}
\sigma_k^- (\boldx) \; = &\; (\mathbf r , e_1 , \dots , e_{k-1} + 1, e_k -1, e_{k+1}, \dots , e_N), \\
\sigma_k^+ (\boldx) \; = &\; (\mathbf r , e_1 , \dots , e_{k-1}, e_k -1, e_{k+1} + 1, \dots , e_N).
\end{align*}
So, with such an interaction, particles tend to give part of their energy to their neighbors at a rate proportional to the value of their energy times the value of $\chi$. 
The instantaneous current is defined through the relation 
\begin{equation*}
L e_k \; = \; \epsilon L'' e_k \; = \; \epsilon \big( j_{k-1,k} - j_{k,k+1} \big) 
\end{equation*}
with 
\begin{equation}\label{current}
j_{k,k+1} \; = \; \chi (r_k, r_{k+1}) (e_k - e_{k+1}).
\end{equation}

\paragraph{Model 2: Exchanging the whole energy.}
Let us further restrict the set of reachable energies to $\{0,1\}$, so that the phase space now writes $\mathsf X = \mathsf M^N \times \{0,1\}^N$.
Let $1 \le k \le N$.
One defines 
\begin{equation*}
L''_k u \; = \; \chi (r_k , r_{k+1}) \big( u\circ \sigma_k - u \big)
\end{equation*}
with 
\begin{equation*}
\sigma_k \mathbf x \; = \; (\mathbf r , e_1 , \dots , e_{k+1}, e_k , \dots , e_N). 
\end{equation*}
So here, near particles may exchange their energy. 
It is checked that the instantaneous current is the same as in the first model. 

Let us stress at this point that the dynamics still should perfectly make sense if energy was allowed to take more than two values. 
However, since energy is only exchanged between particles, there should be $n-1$ conserved quantities if energy was allowed to take $n$ values.
It is not at all clear in that case that an initial density of energy should evolve autonomously in the hydrodynamic regime, 
since an initial macroscopic state should in fact be specified by the value of each conserved quantity. 
We have thus limited ourselves to the case of two reachable energies only, in order to avoid extra complications.

\paragraph{Model 3: Sharing energy.}
Energy is here allowed to take any non negative value, so that the phase space is given by $\mathsf X = \mathsf M^N \times \R_+^N$. 
Let $1 \le k \le N$.
One defines 
\begin{equation*}
L''_k u \; = \; 2\,\chi (r_k , r_{k+1}) \big( T_k - Id \big) u 
\end{equation*}
with 
\begin{equation*}
T_k u \; = \; \int_0^1 u (\boldrr , e_1, \dots , p(e_k + e_{k+1}), (1-p)(e_k + e_{k+1}), \dots , e_N) \, \dd p.
\end{equation*}
The instantaneous current is the same as in the two previous models. 
A similar dynamics is considered in \cite{eck}.

\section{Fluctuation field}\label{sec: Fluctuation field}

The derivation of the heat equation requires to consider an energy profile initially out of equilibrium. 
In this work, like in \cite{oll}\cite{her}, we will however only look at small perturbations out of equilibrium.
To see what small means, 
let us assume that our system is in equilibrium at an inverse temperature $0 < \beta < +\infty$ (see below), 
and let us denote by $\langle e \rangle_\beta$ the average energy of a particle. 
Let $I \subset \T$ be an interval. 
For a typical event in the invariant measure, one has
\begin{equation*}
\sum_{k : k/N \in I} \big( e_k - \langle e \rangle_\beta \big) \; \sim \; \sqrt{N}. 
\end{equation*}
So let
\begin{equation*}
\mathcal E_{t} (I) \; = \; \frac{1}{\sqrt N} \sum_{k: k/N \in I} \big( e_k (t) - \langle e \rangle_\beta \big)
\end{equation*}
be the macroscopic observable we look at.
The random variable $\mathcal E_t (I)$ is expected to fluctuate in a diffusive time-scale. 
So a typical kind of probability that one would like to estimate in the limit $N\rightarrow \infty$ is  
\begin{equation*}
\Proba \big( \mathcal E_{N^2 t} (I) \simeq b \, | \, \mathcal E_{0} (I) \simeq a \big),
\end{equation*}
where $a,b\in\R$, and where  $x \simeq a$ means $x \in [a-\delta, a+ \delta]$ for some small $\delta >0$.
So in fact, we are considering initial profiles that have an excess of energy of order $\sqrt{N}$ in subintervals of $\T$, rather than of order $N$ for a profile truly out of equilibrium

We now need some definitions. 
Let us first define canonical equilibrium measures.
For our three models
\begin{equation*}
\langle u \rangle_\beta \; = \; 
\int_{\mathsf X} u ( \boldrr ,  \bolde) \, \mu_\mathrm U (\dd  \boldrr ) \, \mu_\beta (\dd \mathbf e),
\end{equation*}
where, for $u = u (\bolde)$, one has
\begin{align*}
\mu_\beta (u) \; &= \; 
\frac{1}{Z(\beta)} \sum_{\bolde \in \N^N} \frac{(1/\beta)^{e_1 + \dots + e_N}}{e_1 ! \dots e_N !} \, u (\bolde)
\qquad \text{for Model 1,} \\
\mu_\beta (u) \; &= \; 
\frac{1}{Z(\beta)} \sum_{\bolde \in \{0,1\}^N} \ed^{-\beta (e_1 + \dots + e_N)} \, u(\bolde)
\qquad \text{for Model 2,} \\
\mu_\beta (u) \; &= \; \frac{1}{Z(\beta)} \int_{\R_+^N} \ed^{-\beta (e_1 + \dots + e_N)} \, u (\bolde) \, \dd \bolde
\qquad \text{for Model 3,}
\end{align*}
with $Z(\beta)$ a normalization factor.

Let us next define the fluctuation field $\big( Y^{(N)}_{N^2 t}\big)_{t\ge 0}$. 
For any smooth test function $H$ on $\T$, we define
\begin{equation*}
Y^{(N)}_{N^2t} (H) 
\; = \;
\frac{1}{\sqrt N} \sum_{k=1}^N H(k/N) \big( e_k (N^2t) - \langle e \rangle_\beta \big).
\end{equation*}
Most of the time we will just write $Y$ instead of $Y^{(N)}$.
Formally, $(Y_{N^2t})_{t}$ is as a random process with values in $\mathcal D (\R_+, \mathcal H_{-3})$, 
where $\mathcal D$ denotes the set of cad-lag functions, 
and where $\mathcal H_{-3}$ denotes the Sobolev space of distributions $Z$ on $\T$ with norm $\| Z \|_{-3}^2 = \sum_{k\in \Z} k^{-6} |Z(\mathsf f_k)|^2$,
with $\mathsf f_k (x) = \ed^{2i \pi k x}$. 

Let us finally define the stationary generalized Ornstein-Uhlenbeck process $(\mathrm Y_t)_{t\ge 0} \in \mathcal C (\R_+, \mathcal H_{-3})$, 
where $\mathcal C$ denotes the set of continuous functions, for some diffusion constant $\epsilon  D >0$ and thermal capacity $\sigma^2 > 0$. 
We here have anticipated the fact that, in our applications, the diffusion constant will be of order $\epsilon$.
Given a smooth test function $H$ on $\T$, and for $t\ge 0$, we let $H_t$ be the solution to the heat equation
\begin{equation}\label{evolution heat equation}
H_0 = H, \qquad \partial_t H_t \; = \; \epsilon  D \multiplication \partial^2_x H_t.
\end{equation}
Our definition of $(\mathrm Y_t)_{0 \le t \le T}$ is then as follows. 
First, $\mathrm Y_0(H)$ is a centered Gaussian random variable with variance 
\begin{equation}\label{variance Ornstein Ulhenbeck}
\sigma^2 \int_\T H^2_0 (x) \, \dd x.
\end{equation}
Next, for $t \ge 0$, we have the decomposition
\begin{equation*}
\mathrm Y_t (H) \; = \; \mathrm Y_0 (H_t) +  M_t (H),
\end{equation*}
where $M_t(H)$ is a Gaussian centered martingale, independent of $\mathrm Y_0 (H_t)$, and with variance
\begin{equation*}
2 \epsilon D \sigma^2 \int_0^t \bigg( \int_\T (\partial_x H_s (x))^2\, \dd x \bigg) \, \dd s.
\end{equation*}
Since $\mathrm Y_t$ is a distribution, it holds that $\mathrm Y_t(aG + bH) = a \mathrm Y_t (G) + b Y_t (H)$ for any smooth $G$ and $H$ on $\T$, and $a,b\in \R$. 
Using this property, it is direct to recover the more classical definition of the stationary Ornstein-Uhlenbeck process: $(\mathrm Y_t)_{t\ge 0}$ is the only stationary Gaussian process
with zero mean and covariance 
\begin{equation*}
\Mean \big( \mathrm Y_t (H) \mathrm Y_0 (G) \big)
\; = \; 
\sigma^2 \int_{\R^2} \frac{\ed^{-\frac{(x-y)^2}{4\epsilon t D}}}{\sqrt{4\pi \epsilon t D}} H(x) G(y) \, \dd x \dd y.
\end{equation*}

Our aim is to show that $(Y_{N^2 t})_t$ converges in some sense to $(\mathrm Y_t)_t$ as $N\rightarrow \infty$, with
\begin{equation*}
D \; = \; D(\beta) \; > \; 0
\qquad \text{and} \qquad
\sigma^2 \; = \; \sigma^2(\beta) \; = \; \langle e^2 \rangle_\beta - \langle e \rangle_\beta^2. 
\end{equation*}
We say that a function $D\multiplication\nabla e$ on $\mathsf X$ is gradient if it is of the form 
\begin{equation}\label{gradient function}
D\multiplication \nabla e \; = \; \sum_{k\in\Z_N} D_k \multiplication (e_k - e_{k+1}).
\end{equation}
We then also write $D = \sum_{k\in \Z_N} D_k$.
This quantity depends on the representation \eqref{gradient function} of the gradient function $D \multiplication\nabla e$, which is not unique due to periodic boundary conditions. 
This however will create no trouble in the sequel. 
We will use the following proposition, shown in Section \ref{sec: proof limit fluctuation field}, to derive our results.
The hypotheses are close to optimal for the proof, but could be quite simplified for our applications. 

\begin{Proposition}\label{the: limit fluctuation field}
Let $\epsilon' \ge 0$.
Let $\beta >0$.
Let $\sigma^2 (\beta) = \langle e^2 \rangle_\beta - \langle e \rangle_\beta^2$.
Assume that there are sequences of functions $\big((D\multiplication \nabla e)_N\big)_N$, $(u_N)_N$ and $(g_N)_N$ in $\Lp^2 (\mathsf X, \langle \cdot \rangle_\beta)$, 
that may depend on $\beta$, such that 
\begin{equation*}
- L u_N - \epsilon \, \big(j_{0,1} - (D\multiplication \nabla e)_N \big) \; = \; \epsilon' \, L'' g_N,
\end{equation*}
and that 
\begin{equation*}
\lim_{N\rightarrow \infty} \frac{1}{N^2} \sum_{k\in \Z_N} | \langle \tau_k u_N , u_N \rangle_\beta | \; = \; 0.
\end{equation*}
We moreover assume that, for any $k\in \Z_N$, $(D_k)_N$ converges as $N\rightarrow \infty$,
that the following quantities converge as $N\rightarrow \infty$:
\begin{equation*}
\sum_{k\in{\Z_N}} |(D_k)_N|, 
\qquad
\sum_{k\in\Z_N} |\langle g_N , \tau_k L'' g_N \rangle_\beta|,
\qquad
\sum_{k\in\Z_N} |\langle u_N , j_{k,k+1}\rangle_\beta|, 
\qquad
\sum_{k\in\Z_N} |\langle u_N , \tau_k L'' g_N\rangle_\beta|,
\end{equation*}
and that 
\begin{equation*}
D \; := \; D (\beta) \; := \;  \lim_{N\rightarrow \infty} \sum_{j\in \Z_N} (D_j)_N \; > \; 0.
\end{equation*}

Then, assuming that $X_0$ is distributed according to the equilibrium measure $\langle \cdot \rangle_\beta$, 
the finite dimensional distributions of the process $(Y_{N^2 t})_t$
converge weakly as $N\rightarrow \infty$ to the finite dimensional distributions of a process $(\mathrm Y_t)_t$ such that, for any smooth test function $H$,
\begin{equation*}
\mathrm Y_t (H) \; = \; \mathrm Y_0 (H_t) + M_t(H) + R_t(H),
\end{equation*}
where $H_t$ is given by \eqref{evolution heat equation}.
Here $\mathrm Y_0(H_t)$ is a Gaussian random variable with variance given by \eqref{variance Ornstein Ulhenbeck} with $H_t$ in place of $H_0$,
$M_t(H)$ is a centered Gaussian martingale, independent of $\mathrm Y_0 (H_t)$, which variance at time $t$ is given by 
\begin{equation}\label{limit variance fluctuation field}
2\epsilon \int_0^t \bigg( \int_\T (\partial_x H_s)^2 \, \dd x \bigg) \, \dd s 
\multiplication
\bigg( 
\overline{\chi} \, \sigma^2(\beta)  
- 
\lim_{N\rightarrow \infty}\sum_{k\in\Z_N} \langle u_N, j_{k,k+1} \rangle_\beta + 
\frac{\epsilon'}{\epsilon} \lim_{N\rightarrow \infty}\sum_{k\in\Z_N} \langle u_N,\tau_k L'' g_N \rangle_\beta 
\bigg),
\end{equation}
and $R_t(H)$ is centered with variance at time $t$ bounded by a constant times
\begin{equation*}
(\epsilon')^2 \epsilon^{-1} \multiplication (1 + t^2) \multiplication \lim_{k\in \Z_N}\sum_{k\in \Z_N} | \langle g_N, \tau_k L'' g_N \rangle_\beta |.
\end{equation*} 
\end{Proposition}

\section{Diffusion of energy}\label{sec: Explicit results}

The results below hold for our three models. 
We first obtain the hydrodynamic limit of the field $Y_{N^2 t}$ to the first order in $\epsilon$, with a bound on the variance of the rest term.
As a direct corollary, we obtain a weak coupling limit, where $\epsilon$ is sent to zero after that the diffusive limit is taken. 
\begin{Theorem}\label{the: first order}
Let $D = \overline{\chi}$.
Let us assume that $X_0$ is distributed according to the equilibrium measure $\langle \cdot \rangle_\beta$.
The finite dimensional distributions of the process $(Y_{N^2 t})_t$ converge weakly as $N\rightarrow \infty$ to the finite dimensional distributions of a process $(\mathrm Y_t)_t$ such that 
\begin{equation*}
\mathrm Y_t (H) \; = \; \mathrm Y_0 (H_t) + M_t(H) + R_t(H),
\end{equation*}
where $H_t$ is given by \eqref{evolution heat equation},
where $\mathrm Y_0(H_t)$ is a Gaussian random variable with variance given by \eqref{variance Ornstein Ulhenbeck} with $H_t$ in place of $H_0$,
where $M_t(H)$ is a Gaussian martingale, independent of $\mathrm Y_0(H_t)$, with variance 
\begin{equation*}
2\epsilon D \sigma^2(\beta) \int_0^t \bigg( \int_\T (\partial_x H_s)^2 \, \dd x \bigg) \, \dd s + \mathcal O (\epsilon^2 t)
\qquad \text{(as} \; \; \epsilon \rightarrow 0,\; t\text{ fixed)},
\end{equation*}
and where $R_t(H)$ is centered with a variance that is $\mathcal O (\epsilon^{3} (1 + t^2))$ (as $\epsilon \rightarrow 0$, $t$ fixed).
\end{Theorem}

\Proof
Thanks to \eqref{solution Poisson equation L'}, there exists a function $v\in \L^2(\mathsf X, \langle \cdot \rangle_\beta)$, depending only on $x_0$ and $x_1$, that solves the Poisson equation
\begin{equation*}
- L' v \; = \; j_{0,1} - D \multiplication (e_0 - e_1) \; = \; ( \chi - \overline{\chi} )(r_0,r_1) \multiplication (e_0 - e_1),
\end{equation*}
and such that $\langle v | \mathbf e \rangle = 0$ for every $\mathbf e$.
Defining $\epsilon' = \epsilon^2$, $u_N = \epsilon v$ and $g_N = -v$, one sees that the hypotheses of Proposition \ref{the: limit fluctuation field} are satisfied, 
from which our claim follows. 
$\square$

\begin{Corollary}\label{cor: weak coupling limit}
Let $D = \overline{\chi}$.
Let us assume that $X_0$ is distributed according to the equilibrium measure $\langle \cdot \rangle_\beta$.
Letting first $N \rightarrow \infty$ and then $\epsilon \rightarrow 0$, 
the finite dimensional distributions of the process $(Y_{N^2 \epsilon^{-1}t})_t$ converge weakly to the finite dimensional distributions of 
the stationary Ornstein-Uhlenbeck process with diffusion constant $D$ and thermal capacity $\sigma^2 = \sigma^2 (\beta)$.
\end{Corollary}

\Proof
This follows directly from Theorem \ref{the: first order} thanks to the bound in $\mathcal O (\epsilon^{3} (1 + t^2))$ on the variance of the rest term $R_t$. $\square$

Like in \cite{bri}, we next approximate the internal dynamics of particles by a process independent of energy. 
So we now assume that, for $k \in \Z_N$,
\begin{equation}\label{independent internal dynamics}
L_k' u \; = \;  
\varsigma_k\, \sqrt{\mathsf e_0} \multiplication \partial_{q_k} u 
\, + \,
\sqrt{\mathsf e_0} \multiplication \big( u(\dots , - \varsigma_k, \dots) - u (\dots, \varsigma_k , \dots) \big).
\end{equation}
With such a definition of $L'$, particles evolve now completely independently from each others. 
This in fact is not needed ; 
any operator $L'$, not involving energy and generating a dynamics that quickly relaxes to equilibrium independently of the system size $N$,
should be fine for our purposes. 

Let us see when the diffusion constant obtained with this new definition of $L'$ can be expected to be close to that of our original models. 
For this, let us introduce an extra small parameter $\delta > 0$ in the original models, 
by replacing everywhere the energy $\mathsf e_0 + e_k$ by a new, smaller energy, $\mathsf e_0 + e'_k = \mathsf e_0 + \delta e_k$.
Doing so, the energy of a typical atom of the chain in the Gibbs state will be very close to the positive ground state energy $\mathsf e_0$, if $\delta$ is very small. 
So, for fixed $\epsilon >0$,
we then expect the diffusion constant $\epsilon D =  \epsilon D(\delta)$ to converge to the value given in theorem \ref{the: independent dynamics} below,
valid for any small enough $\epsilon > 0$,
as $\delta \rightarrow 0$.

\begin{Theorem}\label{the: independent dynamics}
Assume that $L'_k$ is given by \eqref{independent internal dynamics} for $k\in \Z_N$, and that $\epsilon > 0$ is small enough.
Assume also that $X_0$ is distributed according to the equilibrium measure $\langle \cdot \rangle_\beta$.
As $N\rightarrow \infty$, the finite dimensional distributions of the field $(Y_{N^2 t})_t$ converge weakly to the finite dimensional distributions of a generalized Ornstein-Uhlenbeck process
with some diffusion constant $\epsilon D >0$ independent of the temperature, and with the thermal capacity $\sigma^2 = \sigma^2 (\beta)$.
\end{Theorem}

\Proof
The proof is made of two steps: 
we first show that Proposition \ref{the: limit fluctuation field} can be applied with $\epsilon' = 0$, 
and we next identify the variance of $M_t (H)$.

\textbf{Step 1.}
Let $n \ge 1$.
We show that the hypotheses of Proposition \ref{the: limit fluctuation field} are satisfied with $\epsilon ' = \epsilon^{n+1}$, 
and that the limit as $N\rightarrow \infty$ of all the quantities appearing in the hypotheses of  Proposition \ref{the: limit fluctuation field}, 
are bounded by a constant that does not depend on $n$.
Let us introduce an extra definition. 
We say that a  function $u$ on $\mathsf X$ is pre-gradient if it is of the form
\begin{equation*}
u (\boldrr, \bolde) \; = \; \sum_{k\in \Z_N} \widehat{u} (\boldrr , k) (e_k - e_{k+1}).
\end{equation*}

Let us first make two observations.
Given a pre-gradient function $f$, the fundamental solution $u$ to the equation $-L' u = f$ is itself pregradient. 
Indeed the coefficients $\widehat{u}(\boldrr , k)$ are the fundamental solutions to the equation $-L' \widehat{u}(\boldrr , k) = \widehat{f}(\boldrr ,k)$.
Second, if a function $u$ is pregradient, so is $L'' u$. 
Indeed,  for the three models, one has 
\begin{equation}\label{expression for L'' u for independent dynamics}
L'' u (\boldrr , \bolde) \; = \; 
\sum_{k\in \Z_N}
\big( 
\widehat{u} (\boldrr , k-1) - 2 \widehat{u} (\boldrr , k) + \widehat{u} (\boldrr,k+1)
\big)
\, \chi (r_k , r_{k+1}) 
\, (e_k - e_{k+1}). 
\end{equation}
So moreover, if $u$ depends on variables $x_a, \dots , x_b$ with $a\le b$, then $L'' u$ depends on variables $x_{a-1} , \dots , x_{b+1}$.

Now, for $N$ large enough for given $n$, the functions $u_N$, $ (D \multiplication \nabla e)_N $ and $g_N$ that we will construct will actually not depend on $N$, 
so that we drop the index $N$.
We set 
\begin{equation}\label{recursive solution Poisson equation}
u \; = \; \sum_{k=1}^n \epsilon^k  u^{(k)},
\qquad
 (D \multiplication \nabla e) \; = \; \sum_{k=1}^n \epsilon^{k-1} \, D^{(k)} \multiplication \nabla e
\qquad \text{and} \qquad 
g \;= \; - u^{(n)},
\end{equation}
with the following recursive definitions of $u^{(k)}$ and $D^{(k)}$ for every $k \ge 1$. 
First 
\begin{equation*}
D^{(1)} \multiplication \nabla e = \langle j_{0,1} | \mathbf e \rangle = \overline{\chi} \, (e_0 - e_1), 
\end{equation*}
and $u^{(1)}$ is the fundamental solution of 
\begin{equation*}
-L' u^{(1)} = j_{0,1} - \langle j_{0,1} | \mathbf e \rangle.
\end{equation*}
The function $u^{(1)}$ is pregradient.
Next, given the pregradient function $u^{(k)}$ for $k \ge 1$, one takes 
\begin{equation*}
D^{(k+1)}\multiplication \nabla e =  \langle L'' u^{(k)} | \bolde \rangle_\beta 
\end{equation*}
This indeed defines a linear gradient since $L'' u^{(k)}$ is pregradient. 
Then one defines $u^{(k+1)}$ to be the fundamental solution to 
\begin{equation*}
-L' u^{(k+1)} = L'' u^{(k)} -  \langle L'' u^{(k)} | \bolde \rangle_\beta.  
\end{equation*}
The function $u^{(k+1)}$ is pregradient. 

The functions $u$, $D\multiplication \nabla e$ and $g$ defined by \eqref{recursive solution Poisson equation} satisfy
\begin{equation}\label{Poisson equation in proof}
- L u - \epsilon (j_{0,1} - D\multiplication \nabla e) \; = \; \epsilon^{n+1} L'' g. 
\end{equation}
Thanks to \eqref{solution Poisson equation L'} and the convergence of geometric series, 
the quantities appearing in the hypotheses of Proposition \ref{the: limit fluctuation field} are then indeed bounded independently of $n$ if $\epsilon >0$ is taken small enough. 
Moreover, since $D^{(1)} = \overline{\chi} > 0$, it follows from \eqref{recursive solution Poisson equation} that $D > 0$ if $\epsilon$ is small enough.

\textbf{Step 2.}
Let $u$, $D\multiplication \nabla e$ and $g$ be given by \eqref{recursive solution Poisson equation}
Let us first give an expression for $D$.
Since $\langle u | \mathbf e \rangle = 0$ for every $\mathbf e$, one has also $\langle L' u | \mathbf e \rangle = 0$ for every $\mathbf e$, 
so that, by \eqref{Poisson equation in proof},
\begin{equation*}
D\multiplication \nabla e
\; = \;
\overline{\chi} (e_0 - e_1) + \langle L'' u | \mathbf e \rangle +  \epsilon^{n} \langle L'' g | \mathbf e \rangle.  
\end{equation*}
The function $L'' u$ is computed by means of \eqref{expression for L'' u for independent dynamics}.
Writing $D \multiplication \nabla e = \sum_{j\in\Z_N} D_j  (e_j - e_{j+1})$, one has 
\begin{equation*}
D
\; = \; 
\sum_{j\in\Z_N} D_j
\; = \; 
\overline{\chi} + 
\sum_{k\in \Z_N} \big\langle \chi (\cdot) \big( \widehat{u} (\cdot , k-1) - 2 \widehat{u} (\cdot , k) + \widehat{u} (\cdot,k+1)\big) \big\rangle
+ \mathcal O (\epsilon^n).
\end{equation*}
Taking on the other hand the expression \eqref{limit variance fluctuation field} for the variance of $M_t(H)$, and using \eqref{expression for L'' u for independent dynamics}
to compute $\lim_{N\rightarrow \infty}\sum_{k\in\Z_N} \langle u_N, j_{k,k+1} \rangle $,
one gets
\begin{equation*}
\sigma^2 D \; = \; \sigma^2 \overline{\chi} 
- 
\lim_{N\rightarrow \infty}\sum_{k\in\Z_N} \langle u_N, j_{k,k+1} \rangle
+ \mathcal O(\epsilon^n). 
\end{equation*}
This shows our claim.
$\square$

\section{Proof of Proposition \ref{the: limit fluctuation field}}\label{sec: proof limit fluctuation field}

We first remind a general formula, that we will use several times.
If $u = u(x,t)$ is a regular enough function on $\mathsf X \times \R_+$, then the process
\begin{equation*}
M_t \; = \; \int_0^t L u (X_s,s) \, \dd s + \int_0^t \partial_t u (X_s,s) \, \dd s + u(X_0,0) - u(X_t,t)
\end{equation*}
is a centered martingale, which variance at equilibrium is given by 
\begin{equation*}
\Mean_\beta \big( M_t \big)^2
\; = \; 
2 \int_0^t \langle u (\cdot ,s), (-L) u (\cdot, s) \rangle_\beta \, \dd s.
\end{equation*}
Let us then make some comments on the notations. 
We will write $\langle \cdot \rangle$ for $\langle \cdot \rangle_\beta$.
For $k\in \Z_N$, we define the centered variables
\begin{equation*}
\eta_k \; = \; e_k - \langle e \rangle.
\end{equation*}
We will write gradient functions as $D\multiplication \nabla \eta$ instead of $D\multiplication \nabla e$.
We will not write explicitly the dependence in $N$ of the functions $u_N$, $g_N$ and $(D\multiplication \nabla \eta)_N$ 
appearing in the hypotheses of Proposition \ref{the: limit fluctuation field}.
We will use the same symbol $\partial_x f$ for both the true derivative of $f$ and its approximative derivative $N\big( f(x+ 1/N) - f(x) \big)$.  

\ProofOf{of Proposition \ref{the: limit fluctuation field}}
Let $H = H(x)$ be the smooth test function, 
and let $G = G(x,s)$ be defined for $0 \le s \le N^2 t$, by 
\begin{equation}\label{equation satisfied by G}
N^2 \partial_t G + \epsilon D \multiplication \partial_x^2 G = 0, 
\qquad 
G(\cdot, N^2 t) = H(\cdot).
\end{equation}
Since $L \eta_k = \epsilon (j_{k-1,k} - j_{k,k+1})$ for $k\in\Z_N$, one has 
\begin{equation*}
\int_0^{N^2 t} \frac{1}{\sqrt N} \sum_{k\in\Z_N} G(k/N,s) L \eta_k \, \dd s
\; = \; 
\int_0^{N^2 t} \frac{1}{\sqrt N} \sum_{k\in\Z_N} G(k/N,s) \epsilon (j_{k-1,k} - j_{k,k+1}) \, \dd s
\end{equation*}
An integration by parts leads thus to
\begin{align*}
Y_{N^2 t} (G (\cdot,N^2 t)) - Y_0 (G (\cdot ,0))
\; = &\; 
- M^{(0)}_{N^2 t}
+ \int_0^{N^2 t} \frac{1}{\sqrt N} \sum_{k\in\Z_N} \partial_t G(k/N,s) \eta_k\, \dd s\\
&\; + \frac{1}{N} \int_0^{N^2 t} \frac{1}{\sqrt N} \sum_{k\in \Z_N} \partial_x G(k/N,s) \epsilon j_{k,k+1} \, \dd s,
\end{align*}
where $M^{(0)}_{N^2 t}$ is a martingale. 
Using then the approximate fluctuation-dissipation decomposition of the current stated in the hypotheses of Proposition \ref{the: limit fluctuation field}, one gets
\begin{align}
Y_{N^2 t} (G(\cdot,N^2 t)) - Y_0 (G(\cdot ,0))
\; = &\;
- M^{(0)}_{N^2 t}
+ \int_0^{N^2 t} \frac{1}{\sqrt N} \sum_{k\in\Z_N} \partial_t G(k/N,s) \eta_k\,  \dd s 
\label{proof term 1}\\
&\; - \frac{1}{N} \int_0^{N^2 t} \frac{L}{\sqrt N} \sum_{k\in \Z_N} \partial_x G(k/N,s) \tau_k u \, \dd s 
\label{proof term 2}\\
&\; + \frac{\epsilon}{N} \int_0^{N^2 t} \frac{1}{\sqrt N} \sum_{k\in\Z_N}  \partial_x G(k/N,s)\tau_k D\multiplication \nabla \eta \, \dd s 
\label{proof term 3}\\
&\; - \frac{\epsilon'}{N}\int_0^{N^2 t} \frac{L''}{\sqrt N} \sum_{k\in \Z_N} \partial_x G(k/N,s) \tau_k g \, \dd s
\label{proof term 4}.
\end{align}

We now proceed to the analysis of the last three terms in the right hand side of this equation. 
Let us start with \eqref{proof term 2}.
Since $\partial_t \partial_x H(k/N,s) = -(\epsilon D/N^2)\partial_x^3 H(k/N,s)$, there exists a martingale $M^{(1)}_{N^2 t}$ such that 
\begin{align*}
\eqref{proof term 2} 
\; = &\; 
 - M_{N^2 t}^{(1)}
- \frac{\epsilon D}{N^3}\int_0^{N^2 t} \frac{1}{\sqrt N} \sum_{k\in \Z_N} \partial_x^3 G(k/N,s) \, \tau_k u \, \dd s \\
&\; + \frac{1}{N}\frac{1}{\sqrt N}\sum_{k\in\Z_N} \partial_x G (k/N,N^2 t) \, \big( \tau_k u (X_0) - \tau_k u (X_{N^2 t}) \big)\\
=: &\; - M_{N^2 t}^{(1)} + \rho^{(0)}_{N^2 t}
\end{align*}
The hypothesis that $\sum_{k\in\Z_N} |\langle \tau_k u , u\rangle_\beta|$ remains bounded as $N\rightarrow \infty$, 
ensures that $\Mean (\rho^{(0)}_{N^2 t})^2 \rightarrow 0$ as $N\rightarrow \infty$. 
Let us then look at \eqref{proof term 3}.
Reminding that $D = \lim_{N\rightarrow \infty}\sum_{k\in\Z_N}D_k$, an integration by parts yields
\begin{align*}
\eqref{proof term 3}
\; = &\;
\frac{\epsilon D}{N^2} \int_0^{N^2 t} Y_s (\partial_x^2 G(\cdot, s)) \, \dd s\\
& \; + 
\sum_{j} \frac{\epsilon D_j}{N^2} \int_0^{N^2 t} \frac{1}{\sqrt N} \sum_k \big( \partial_x^2 G(k/N,s) - \partial_x^2 G((k+j + 1)/N,s) \big) \eta_{k+j+1}\, \dd s \\
=: &\;
\frac{\epsilon D}{N^2} \int_0^{N^2 t} Y_s (\partial_x^2 H(\cdot, s)) \, \dd s + \rho^{(1)}_{N^2 t}.
\end{align*}

Let us show that $\Mean (\rho^{(1)}_{N^2 t})^2 \rightarrow 0$ as $N\rightarrow \infty$. 
First, by Jensen's inequality, 
\begin{equation*}
\Mean  (\rho^{(1)}_{N^2 t})^2 
\; \le \;
\frac{t^2}{N^2 t} \int_0^{N^2 t} \Mean \Big( 
\sum_{j} \epsilon D_j \frac{1}{\sqrt N} \sum_k \big( \partial_x^2 G(k/N,s) - \partial_x^2 G((k+j + 1)/N,s) \big) \eta_{k+j+1} (s)
\Big)^2
\, \dd s.
\end{equation*}
At this point, we split the sum over $j$ as
\begin{equation*}
\sum_j (\dots) \; = \; \sum_{j:|j| \le \sqrt N} (\dots) \; + \; \sum_{j: |j| > \sqrt N} (\dots), 
\end{equation*}
and we separately show that the variance of each of these terms, at a fixed time $s \in [0,N^2 t]$, converges to 0, which will establish that $\Mean (\rho^{(1)}_{N^2 t})^2 \rightarrow 0$.
Since  $\sum_{j\in\Z_N}|D_j|$ remains bounded as $N\rightarrow \infty$, it holds that 
\begin{equation}\label{a variance somewhere in a proof...}
\Mean \Big(\sum_{j:|j| \le \sqrt N} (\dots)\Big)^2 \; \le \; \mathrm C \max_{j:|j|\le \sqrt N} \Mean (\dots)^2
\end{equation}
for some constant $\mathrm C < + \infty$.
But then
\begin{equation*}
\big| \partial_x^2 G(k/N,s) - \partial_x^2 G((k+j + 1)/N,s) \big| \; = \;  \mathcal O (|j+1|/N) \; = \; \mathcal O (1/ \sqrt N), 
\end{equation*}
so that the right hand side of \eqref{a variance somewhere in a proof...} is seen to converge to 0. 
For the second term, we have
\begin{equation*}
\Mean \Big( \sum_{j: |j| > \sqrt N} (\dots) \Big)^2 \; \le \; 
\Big( \sum_{j: |j|\ge \sqrt N} D_j\Big)^2
\Mean \Big( \frac{\epsilon}{\sqrt N} \sum_k \big( \partial_x^2 G(k/N,s) - \partial_x^2 G((k+j + 1)/N,s) \big) \eta_{k+j+1} (s) \Big)^2.
\end{equation*}
We remind that, in this expression, $D_j$ is written for $(D_j)_N$.
Since, by hypothesis, both $(D_j)_N$ and $\sum_{l\in\Z_N}|D_l|$ converge as $N\rightarrow \infty$, for any given $j\in \Z_N$,
we conclude that the first factor in the right hand side of this inequality converges to 0 as $N\rightarrow \infty$. 
Since the second factor stays bounded as $N\rightarrow \infty$, we deduce that the left hand side converges to 0 as $N\rightarrow \infty$.

We finally handle the less usual term \eqref{proof term 4}, which is a centered process.
It is convenient to define
\begin{equation*}
F_s \; = \; \frac{1}{\sqrt N}\sum_{k\in\Z_N} \partial_x G(k/N,s) \, \tau_k g, 
\qquad
\tilde F_s \; = \; \frac{1}{\sqrt N}\sum_{k\in\Z_N} \partial^3_x G(k/N,s) \, \tau_k g
\end{equation*}
For $z>0$ one then defines $V_{z,s}$ and $\tilde V_{z,s}$ as respective unique solutions to the equations
\begin{equation}\label{Poisson equation with z}
(z-L) V_{z,s} \; = \; - L'' F_s, 
\qquad
(z-L) \tilde V_{z,s} \; = \; - L'' \tilde F_s.
\end{equation}
So, defining a centered martingale $M^{(2)}_{N^2 t}$, one has
\begin{align*}
\eqref{proof term 4}
\; = &\; 
\epsilon' \frac{1}{N} \int_0^{N^2 t} (-L'') G_s (X_s) \, \dd s \\
\; = &\; 
\frac{\epsilon' z}{N} \int_0^{N^2 t} V_{z,s} (X_s) \, \dd s
- \frac{\epsilon'}{N} \int_0^{N^2} L V_{z,s}(X_s) \, \dd s\\
\; = &\; 
\frac{\epsilon' z}{N} \int_0^{N^2 t} V_{z,s} (X_s) \, \dd s
+ \frac{\epsilon'}{N} \int_0^{N^2 t} \partial_t V_{z,s} (X_s) \, \dd s
- \frac{\epsilon'}{N} M_{N^2 t}^{(2)} + \frac{\epsilon'}{N} \big( V_{z,0}(X_0) - V_{z,N^2 t}(X_{N^2 t}) \big)
\end{align*}
Therefore, taking $z = 1/N^2$, one finds a constant $\mathrm C < +\infty$ such that
\begin{align*}
\Mean_\beta \big(\eqref{proof term 4}^2 \big)
\; \le &\;\mathrm C\,(\epsilon')^{2} 
\bigg(
t^2  \multiplication \frac{1}{N^2 t}\int_0^{N^2 t} z \langle V_{z,s}^2 \rangle \, \dd s + t^2 \multiplication \frac{1}{N^2 t}\int_0^{N^2 t} z \langle \tilde V_{z,s}^2 \rangle \, \dd s \\
&\; +
  t \multiplication \frac{1}{N^2 t}\int_0^{N^2 t} \langle V_{z,s}, (-L) V_{z,s} \rangle \, \dd s
+   z \langle V_{z,0}^2 \rangle + z \langle V_{z,N^2 t}^2 \rangle
\bigg).
\end{align*}
 But one has the bounds
\begin{equation}\label{proof a bound}
z \langle V_{z,s}^2 \rangle \; \le \; \epsilon^{-1} \langle F_s , (-L'')F_s \rangle, 
\qquad
\langle V_{z,s}, (-L)V_{z,s} \rangle \; \le \; \epsilon^{-1} \langle F_s , (-L'')F_s \rangle,
\end{equation}
as well as the same bounds with the tilted quantities. 
Indeed, \eqref{Poisson equation with z} implies
\begin{align*}
z \langle V_{z,s}^2 \rangle + \langle V_{z,s} \multiplication (-L) V_{z,s} \rangle
\; = &\; 
\langle V_{z,s} \multiplication (-L'') F_s \rangle
\; \le \; 
\langle F_s \multiplication (-L'') F_s \rangle^{1/2} \langle V_{z,s} \multiplication (-L'') V_{z,s} \rangle^{1/2} \\
\; \le &\; 
\epsilon^{-1/2}\, \langle F_s \multiplication (-L'') F_s \rangle^{1/2} \langle V_{z,s} \multiplication (-\epsilon \, L'') V_{z,s} \rangle^{1/2} \\
\; \le &\; 
\epsilon^{-1/2}\, \langle F_s \multiplication (-L'') F_s \rangle^{1/2} \langle V_{z,s} \multiplication (-L) V_{z,s} \rangle^{1/2},
\end{align*}
from where \eqref{proof a bound} is derived. 
So finally we obtain
\begin{equation}\label{bound on the variance of R}
\Mean_\beta \big(\eqref{proof term 4}^2 \big) 
\; \le \;
\mathrm C \,(\epsilon')^2 \epsilon^{-1} (1+ t^2) \lim_{N\rightarrow \infty}\sum_{k\in\Z_N} |\langle g, L'' \tau_k g \rangle_\beta|.
\end{equation}

So at this point, let us define
\begin{equation*}
R_{N^2 t} \; = \; \eqref{proof term 4} 
\qquad \text{and} \qquad
M_{N^2 t} \; = \; - M_{N^2 t}^{(0)} - M_{N^2 t}^{(1)}.
\end{equation*} 
Taking into account that $G$ satisfies \eqref{equation satisfied by G}, one has so far obtained
\begin{equation*}
Y_{N^2 t} (H) 
\; = \; 
Y_0(G(\cdot ,0)) + M_{N^2 t} + R_{N^2 t} + \rho_{N^2 t}^{(0)} + \rho^{(1)}_{N^2 t}.
\end{equation*}
It holds that $G(\cdot , 0) = H_t (\cdot)$, that $\Mean_\beta (R^2_{N^2 t})$ is bounded by \eqref{bound on the variance of R} 
and that both $\rho_{N^2 t}^{(0)}$ and $\rho_{N^2 t}^{(1)}$ converge to 0 in $\Lp^2(\Proba_\beta)$. 
Therefore, if we establish that
\begin{multline}
\lim_{N\rightarrow \infty}\Mean_{\beta} (  M_{N^2 t}^2 ) 
\; = \; \\
2\epsilon
\lim_{N\rightarrow \infty}
\frac{1}{N^2}\int_0^{N^2 t}\hspace{-0.1cm}\bigg( \int (\partial_x G)^2 (x,s) \, \dd x \bigg) \dd s
\multiplication
\Big(
\overline{\chi}\sigma^2
-
\sum_{k\in\Z_N} \langle u , j_{k,k+1}  \rangle
+
\frac{\epsilon'}{\epsilon}
\sum_{k\in \Z_N} \langle u, \tau_k L''g \rangle 
\Big),
\label{convergence of the variance}
\end{multline} 
then we will be able to conclude that $ M_{N^2 t}$ converges as a process to a centered Gaussian martingale, 
independent of $Y_0(H_t)$, with the variance given by \eqref{limit variance fluctuation field} (see \cite{kom} or also \cite{her}). 
It thus remains to show \eqref{convergence of the variance} to conclude the proof. 

Writing $Y_s$ for $Y_s (H(\cdot , s))$ and putting 
\begin{equation*}
U_s \; = \; \frac{1}{N}\frac{1}{\sqrt{N}} \sum_{k\in\Z_N} \partial_x H (k/N,s) \tau_k u , 
\end{equation*}
the variance of $M_{N^2 t}$ is given by 
\begin{align*}
\Mean (M_{N^2 t})^2 
\; = &\; 2 \int_0^{N^2 t} \langle Y_s + U_s , (-L) (Y_s + U_s) \rangle \, \dd s \\
\; = &\; 
2\epsilon \int_0^{N^2 t} \langle Y_s , (-L'')Y_s \rangle \, \dd s
+ 4 \epsilon \int_0^{N^2 t} \langle U_s, (-L'') Y_s \rangle \, \dd s
+ 2 \int_0^{N^2 t} \langle U_s , (-L)U_s \rangle \, \dd s.
\end{align*}
One first readily computes that, since $\sum_{k\in\Z_N}|\langle u , j_{k,k+1}\rangle|$ remains bounded as $N\rightarrow \infty$,
\begin{align*}
2\epsilon \int_0^{N^2 t} \langle Y_s , (-L'')Y_s \rangle \, \dd s
\; \rightarrow \; &
2 \epsilon \overline{\chi} \sigma^2 \multiplication \lim_{N\rightarrow \infty}\frac{1}{N^2}\int_0^{N^2 t}\bigg( \int (\partial_x G)^2 (x,s) \, \dd x \bigg)\, \dd s,\\
4 \epsilon \int_0^{N^2 t} \langle U_s, (-L'') Y_s \rangle \, \dd s
\; \rightarrow \;&
- 4 \epsilon \lim_{N\rightarrow \infty} \frac{1}{N^2} \int_0^{N^2t } \bigg( \int (\partial_x G)^2 (x,s) \, \dd x\bigg) \, \dd s 
\multiplication \sum_{k\in\Z_N} \langle u , j_{k,k+1} \rangle.
\end{align*}
Then, one uses the approximate  fluctuation-dissipation decomposition of the current stated as hypothesis in Proposition \ref{the: limit fluctuation field} to obtain 
\begin{equation*}
-L U_s 
\; = \;
 \frac{\epsilon}{N^{3/2}} \sum_{k\in\Z_N} \partial_x G (k/N,s) j_{k,k+1} - 
\frac{\epsilon}{N^{3/2}}  \sum_{k\in\Z_N} \partial_x G (k/N,s) \,\tau_k D\multiplication \nabla \eta
+ \frac{\epsilon'}{N^{3/2}} \sum_{k\in\Z_N} \partial_x G(k/N,s)\, \tau_k L'' g.
\end{equation*}
The second term in the right hand side of this equation will not contribute since it is a gradient. 
Taking into account that both $\sum_{k\in\Z_N}|\langle u , j_{k,k+1}\rangle|$ and $\sum_{k\in\Z_N}|\langle u , \tau_k L'' g\rangle|$ remain bounded as $N\rightarrow \infty$,
a computation furnishes  
\begin{equation*}
2 \int_0^{N^2 t} \langle U_s , (-L)U_s \rangle \, \dd s
\; \rightarrow \; 
2 \lim_{N\rightarrow \infty}\frac{1}{N^2} \int_0^{N^2 t} \bigg( \int (\partial_x G)^2 (x,s) \dd x \bigg) \, \dd s
\multiplication 
\Big(
\epsilon \sum_{k\in\Z_N} \langle u , j_{k,k+1} \rangle + \epsilon' \sum_{k\in \Z_N} \langle u, \tau_k L'' g \rangle
\Big).
\end{equation*}
So \eqref{convergence of the variance} is shown. $\square$

\noindent 
\textbf{Acknowledgements.}
I am grateful to L.-S. Young for her kind invitation at the Courant Institute, where this work initiated ; 
I am indebted to her for the time she shared with me discussing many ideas related to this manuscript.
It is also a pleasure for me to thank S. Olla, F. Simenhaus and G. Stoltz for illuminating discussions.   
I thank the European Advanced Grant Macroscopic Laws and Dynamical Systems (MALADY) (ERC AdG 246953) for financial support.


\end{document}